\documentclass[dvips,aap]{imsart}

\RequirePackage[OT1]{fontenc}
\RequirePackage{amsthm,amsmath,natbib}
\RequirePackage[colorlinks,citecolor=blue,urlcolor=blue]{hyperref}
\RequirePackage{hypernat}
\usepackage{epsfig}
\usepackage{graphicx}
\usepackage{psfrag}
\usepackage{multirow}
\usepackage{array}
\usepackage{amsfonts}

\arxiv{}

\startlocaldefs
\numberwithin{equation}{section}
\theoremstyle{plain}
\newtheorem{thm}{Theorem}[section]

\newtheorem{lemma}{Lemma}[section]
\newtheorem{prop}{Proposition}[section]

\newcommand{\esp}{\mathbb{E}}
\newcommand{\proba}{\mathbb{P}}

\newcommand{\Num}{N_U(\mot)}

\newcommand{\mot}{\mathbf{m}}
\endlocaldefs

\begin{document}

\begin{frontmatter}
\title{Detecting Local  Network Motifs}
\runtitle{Detecting Local Network Motifs}

\begin{aug}
\author{\fnms{Etienne} \snm{Birmel\'e}\thanksref{t1}\ead[label=e1]{etienne.birmele@genopole.cnrs.fr}},
\ead[label=u1,url]{http://stat.genopole.cnrs.fr}

\thankstext{t1}{This work has been supported by the CNRS}
\thankstext{t2}{This work has been supported by the French Agence Nationale de le Recherche under grant NeMo ANR-08-BLAN-0304-01}
\runauthor{E. Birmel\'e}

\affiliation{Laboratoire Statistique et G\'enome, Universit\'e d'Evry, \\
 UMR CNRS 8071, INRA 1152}

\address{Laboratoire Statistique et G\'enome, Tour Evry 2\\
523 place des Terrasses de l'Agora, 91000 Evry, France\\
\printead{e1}\\
}

\end{aug}

\begin{abstract}

Studying the topology of so-called {\em real networks}, that is networks 
obtained from sociological or biological data for instance, has become a major field of
interest in the last decade. One way to deal with it is to consider that networks are 
built from small functional units called {\em motifs}, which can be found by looking for 
small subgraphs whose numbers of occurrences  in the whole network are
surprisingly high. 
In this article, we propose to define motifs through a local over-representation in the network and 
develop a statistic  to detect them 
without relying on simulations.
We then illustrate the performance of  our procedure on simulated and real data, recovering already known biologically relevant  
motifs. Moreover, we explain how our method gives some information about the respective roles of the vertices in a motif.
\end{abstract}



\end{frontmatter}

\section{Introduction}\label{introsect}

One way to reach a better understanding of the structure of networks is to summarize part of the information in
the counts of small subgraphs. That method is used for decades in social network studies, for example via the triad censuses \citep{WS98}. 
More recent work indicates that biological networks show recurrent small patterns, 
called {\em network motifs} and introduced by \cite{MS02}. They can be thought of as small units of
given function from which the networks are built. For instance, \cite{Al07}
describes the regulation role in transcriptional networks of a pattern of 
three vertices called the feed-forward loop. It is therefore quite natural to ask which are the patterns  that are over-represented in a given network.
\medskip

Looking for over-representation requires a null model to compare the observed network with. The most popular model is the stub-rewiring model introduced by \cite{MS02}, which is used in several methods for motif detection including those of \cite{KIM04, WR06, KAE09}, the method based on graph alignments of \cite{BL04} and the method devoted to labeled graphs of \cite{BNC08}.
It is a model requiring the generation of a large number of graphs whose nodes have the same in and out-degrees as the observed network. 
However, \cite{AFB04} point out that  this method does not take into account the preferential links between some vertices and the high local density, which are two major features of biological networks. They also show that the use of the stub-rewiring model and of a Z-score to detect motifs  may lead to false positives. 
 
Another way to define the null model is to consider a random graph model defined by a probability distribution. Litterature about random graph models and their suitability to real networks is abundant (see e.g. \citealp{CL06}). Among the existing models,  mixture models play an important role as they allow  different link probabilities between classes of vertices and thus model the heterogeneity of connection patterns. Moreover, mean and variance calculation for the pattern counts are tractable under such models, as shown by \cite{PDK08}.
\smallskip

It is important that motifs are defined conditionally on subpatterns occurrences, as pointed out by \citeauthor{MS02}. Indeed, a pattern may appear as over-represented because it contains an over-represented subpattern, which is in fact the biologically relevant structure. 
This conditioning issue is also taken into account by \citeauthor{BNC08} in the context of labeled graph. Nevertheless, in both cases, the real network is compared to graphs generated by the stub-rewiring procedure. Therefore, to study the patterns of size $k$, it is necessary to generate a large  number of graphs with the same number of each type of subgraphs of size ranging between $2$ and $k-1$ as in the observed network. In practice, only the cases $k=3$ and $k=4$ are implemented to our knowledge \citep{MS02}. 
\smallskip

Finally, \cite{DBBO04} show that the motifs found in the Yeast transcriptional regulatory network aggregate, that is they highly concentrate in some regions of the network, indicating that biologically meaningful mechanisms may not be spread uniformly in the network. Therefore, it seems natural to look for {\em local} over-representation of patterns. 

That phenomenon is also highlighted in the more biologically driven work by \cite{ZKW05}, who suggest to look for motif themes rather than  motifs. They define themes as {\em recurring higher-order interconnection patterns that encompass multiple occurrences of network motifs} and show their biological relevance in the different networks associated to Yeast. 
In other words, themes are patterns corresponding to several occurrences of a motif that share some of their nodes.
\medskip

  The major contribution of this paper is to propose a definition of a local motif based on the themes of \citeauthor{ZKW05}, and a procedure to detect the local motifs of fixed size $k$ in a network.  This procedure builds upon earlier works in motif detection, being to our knowledge the first approach taking into account the conditioning on a subpattern without size limitation on the considered pattern, as well as the local character of patterns.
Moreover, it makes no assumption on the law of the pattern counts.  It is composed of four main steps:
\begin{itemize}
  \item Inference of the parameters of the null model. We consider a model in which each node belongs to a fixed class and each edge is drawn independently from the others under a Bernoulli law whose parameter depends only on the classes of its endvertices. 
  \item Enumeration and localization  of all patterns of size $k$ present in the studied network and of their subpatterns.
  \item Assignment of a $p$-value to each pair (pattern; subpattern) present in the network for testing local over-representation. The key idea of that assignment is to show that the distribution of the number of local occurrences of a pattern is close to a Poisson distribution, allowing us to bound the exact $p$-value. 
  \item A filtering procedure which ensures that every emergent local motif conveys some novel information about the network structure when compared to its subpatterns.
\end{itemize}
\medskip

We define precisely the notion of local over-representation and detail the four steps of our procedure in Section~\ref{method}. As the obtained $p$-value is in fact an upper bound of the exact one, we investigate the tightness of that bound in Section~\ref{lowerboundsection}. We then show results on both simulated and real data  in Section~\ref{results}.


\section{Methods}\label{method}
  
\subsection{Local network motifs}

   We consider a network $G$ of interest on $n$ vertices. In this work, we consider directed graphs, with possible loops and opposite edges, but all the results can easily be extended to undirected graphs. 

  A {\em pattern} $\mot$ of size $k$ is a directed graph on $k$ vertices, from which we want to know if it is locally over-represented in $G$. 
  As a convention, we will denote by $(a,b,\dots)$ the vertices of $\mot$ and by $(u,v,\ldots)$ those of $G$.

\begin{figure} [htb] 
\begin{center}
\psfrag{a}{$a$}
\psfrag{b}{$b$}
\psfrag{c}{$c$}
\psfrag{d}{$d$}
\psfrag{aa}{$\{a\},$}
\psfrag{bb}{$\{b\},$}
\psfrag{cc}{$\{c\},$}
\psfrag{ab}{$\{a,b\},$}
\psfrag{cd}{$\{c,d\},$}
\psfrag{dc}{Subpatterns}
\psfrag{m}{Pattern}
\psfrag{draw}{Synthetic drawing}
\psfrag{ffl}{Feed-forward loop}
\psfrag{bf}{Bi-fan}
\epsfig{file=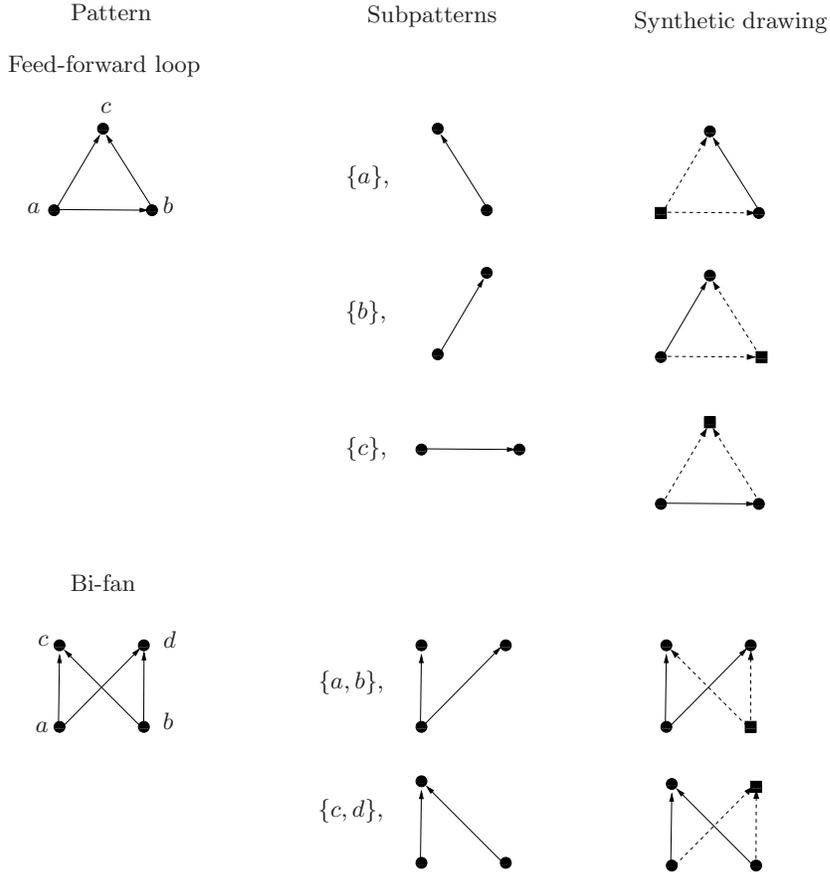, width=.8\textwidth}
\end{center}
\caption{ The feed-forward loop and bi-fan patterns with the list of their subpatterns. The last column shows how we represent a pattern and one of its subpatterns in a single drawing.}
\label{delclasses}
\end{figure}

  An automorphism of $\mot$ is a permutation $\phi$ of its vertices such that, for every pair $(a,b)$ of vertices, $\overrightarrow{\phi(a)\phi(b)}$ is an edge if and only if $\overrightarrow{ab}$ is an edge. Let $\mathcal{R}$ be the relation defined by $a\mathcal{R}b$ if $b$ is the image of $a$ by an automorphism. $\mathcal{R}$ is an equivalence relation on the vertices of $\mot$ and those vertices can therefore be partitioned  into equivalence classes, which we call {\em deletion classes}. For example, in the bi-fan pattern shown in Figure~\ref{delclasses}, the permutation exchanging $a$ with $b$ and $c$ with $d$ is an isomorphism. However, $a$ and $c$ are not equivalent as they have different outdegrees. Thus the bi-fan has two deletion classes which are $\{a,b\}$ and $\{c,d\}$.

Let $(C_1,\ldots,C_K)$ be the deletion classes of $\mot$ and $(i_1,\ldots,i_K)$ their respective sizes. A {\em position} $U$ in a network $G$ for $\mot$ will then denote a list $(V_1,\ldots,V_K)$ of disjoint sets of vertices of $G$ with respective sizes $(i_1,\ldots,i_K)$. That position is an {\em occurrence} of $\mot$ in $G$ if the subgraph of $G$ induced by the vertices of $U$ is isomorphic to $\mot$.  
Writing a position as a list of sets of vertices ensures to count every occurrence of a pattern only once. However, for clarity, we will write positions as lists of vertices throughout the article.
\smallskip

  A {\em subpattern of $\mot$} denotes a pair $(C, \mot')$, where $C$ is a deletion class of $\mot$ and $\mot'$ the pattern on $k-1$ vertices obtained by deleting any vertex of $C$. As all vertices in $C$ are isomorphic, their respective deletion lead to isomorphic subgraphs and the notion of subpattern is thus well defined.

However, the opposite is not true, that is the subgraphs of $\mot$ obtained by deleting a vertex $a$ or a vertex $b$ may be isomorphic while $a$ and $b$ do not belong to the same deletion class. Consider for example the feed-forward loop shown in Figure~\ref{delclasses}.
Deleting any of its three vertices leads to a single edge but all the vertices belong to different deletion classes as they 
are not topologically equivalent (they have for instance different out-degrees).

 In the following, we adopt the graphical convention shown in the last column of Figure~\ref{delclasses} to draw at the same time a pattern $\mot$ and one of its subpatterns $(C,\mot')$. The whole graph represents $\mot$ and the squared vertex whose adjacent edges are dotted is a vertex of $C$. The pattern $\mot'$ is thus obtained by deleting that vertex.

\begin{figure} [htb]
\begin{center}
\psfrag{m1}{$\mot_1$}
\psfrag{m2}{$\mot_2$}
\psfrag{384}{$PDR1$}
\psfrag{388}{$PDR3$}
\psfrag{274}{$PDR5$}
\psfrag{281}{$IPT1$}
\psfrag{300}{$HXT9$}
\psfrag{389}{$HXT11$}
\psfrag{523}{$SNQ2$}
\psfrag{665}{$YOR1$}
\epsfig{file=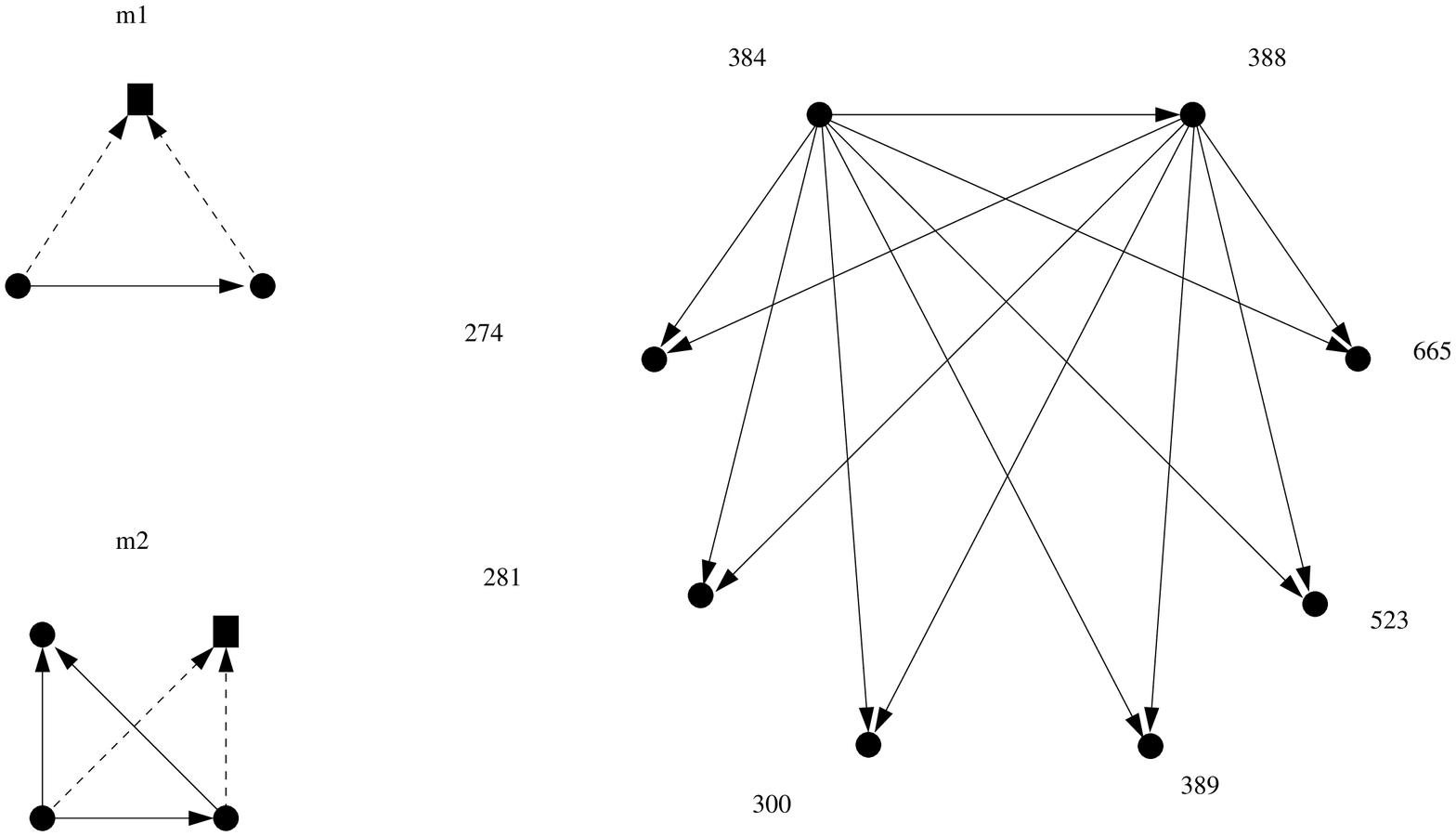, width=.6\textwidth}
\end{center}
\caption{The shown subnetwork of the Yeast regulation network is a $(\mot_1,C_1)$-theme of order 6 at position $(\{PDR1,PDR3\})$ and a $(\mot_2,C_2)$-theme of order 5 at position $(\{PDR1,PDR3\},\{PDR5\})$. $C_1$ and $C_2$ are the respective deletion classes of the squared vertices.}
\label{themeexample}
\end{figure}

  Let $\mot$ be a pattern and $(C,\mot')$ one of its subpatterns. An occurrence $U$ of $\mot$ in $G$ is an {\em extension} of an occurrence $U'$ of $\mot'$ if the vertex set of $U'$ is a subset of the vertex set of $U$ .
  We define the {\em $(\mot,C$)-theme on $U'$} as the subgraph of the network induced by the occurrence of $\mot'$ at $U'$ and all its extensions. The number of those extensions will be the {\em order} of the theme (see Figure~\ref{themeexample} for an illustration).

  We define a {\em potential local motif} as a pattern which is locally over-represented with respect to at least one of its subpatterns. In other words, $\mot$ is a potential local motif with respect to $(C,\mot')$ if there  exist an $(\mot,C)$-theme whose order is significantly higher than the expected order in a random model to be specified in Section \ref{randommodelsection}. 
Note that the local over-representation is different from the global one. Indeed, a pattern having a high number of disjoint occurrences may be globally over-represented without being a motif according to our definition. On the other hand, a pattern  may be locally over-represented without having a large number of occurrences in the whole network.

    Finally, a potential local motif is a {\em local motif} if the information it conveys is not redundant with a smaller local motif, that is if it is not filtered out by the procedure to be described in subsection~\ref{motifselection}.

\subsection{The random graph model}\label{randommodelsection}

 The random generation model we consider is based on blockmodels \citep{WB76}, with a fixed and known class  for each vertex and random edges. It depends on a $4$-tuple of 
parameters $(n,Q,\mathbf{Z},\mathbf{\Pi)}$, where:
 \begin{itemize}
   \item $n$ is the number of vertices,
   \item $Q$ is the number of classes of the model,
   \item $\mathbf{Z}\in \{1,\ldots, Q\}^n$ is a vector giving the class of each vertex, 
   \item $\mathbf{\Pi}$ is a $Q\times Q$ connectivity matrix. The coefficient $\Pi_{ql} \in [0,1]$ of that matrix indicates how likely a vertex of class $q$ and a vertex of class $l$ are linked by an edge.   
\end{itemize}

  Under this model, all the edges of the random graph are drawn independently under Bernoulli laws: denoting by $X_{uv}$ the indicator variable of the edge between vertices $u$ and $v$,

$$  X_{uv} \sim \mathcal{B}(\Pi_{Z(u),Z(v)}). $$  

   Such a model belongs to the family of blockmodels, which are widely used to describe real network data \citep{NS01}. It has at least two main advantages. First, it takes into account the preferential attachment process between several groups of nodes in the network; second, as all the edges are independent, calculations remain tractable.  

  Note that it is not the classical random graph model introduced by Erd\H{o}s and R\'enyi (1959) as the edge probabilities are non uniform (unless $Q=1$). It is neither a mixture model as the classes of the vertices are not random. Adding randomness on the classes would induce correlations between the edges and therefore invalidate the Poisson approximation of Section~\ref{localsection}. 

 Nevertheless, the vertex classes and the connectivity matrix $\Pi$ may be inferred using estimation algorithms for graph mixtures \citep{NS01,DPR08,HW08,LBA08}. The graph mixture models, when applied for $Q$ components in the mixture, infer for each vertex a vector of length $Q$ giving the probability for that vertex to belong to each class, as well as a connectivity matrix. We choose here to assign each vertex to its more probable class, and thus deal with the the observed network as if the groups were known and fixed.  

Our general random graph framework also contains another widely used model where $u$ and $v$  are linked with probability proportional to the product of their respective observed degrees. Under that model, which we will call {\em Expected Degree},   the expected degree of each vertex is almost equal to its observed degree \citep{MSB06}. However, under this model, nodes are in the same class if and only if they have the same in- and outdegrees. Therefore, the number of classes may be large on real networks, having a deep impact on the running time of our motif detection procedure.

Overlapping classes can also be taken into account by using for instance the {\em Mixed Membership Stochastic Blockmodel}  from \cite{ABF08} or the model {\em Overlapping Stochastic Blockmodel} from \cite{LBA09}.

\subsection{Local over-representation}\label{localsection}

Consider a pattern $\mot$ of size $k$, a subpattern $(C, \mot')$ of $\mot$ and a set $U$ of $k-1$ vertices in some graph $G$. If $U$ corresponds to an occurrence of $\mot'$, we write $G[U]\sim \mot'$. 
Let
$N_U(\mot)$ denote the order of the $(\mot, C)$-theme located at $U$. If $U$ does not correspond to an occurrence of
$\mot'$, we set $N_U(\mot)=0$.

We define $\lambda_U(\mot) = \esp(N_U(\mot)|G[U]\sim \mot')$ and $\Delta_U(\mot)$ as the normalized quantity 
$$\Delta_U(\mot)=\frac{N_U(\mot)-\lambda_U(\mot)}{\lambda_U(\mot)} .$$

 Looking for themes whose order is much larger than expected under our model is then equivalent to look for values of $\Delta_U(\mot)$ significantly larger than $1$.

In the following, we will omit the reference to $\mot$ when there is no ambiguity. 
\medskip

Let us consider a set $U$ corresponding to an occurrence of $\mot'$. 
For each vertex $v\notin U$, we denote by $I_U^v$ the indicator random variable which is equal to $1$ if adding $v$ to $U$ yields an occurrence of $\mot$ in $G$.
Let $p_U^v$ be the mean value of $I_U^v$. Then $\lambda_U = \sum_{v\notin U} p_U^v$ and those quantities can easily be deduced from the parameters of the random graph model.

As the indicator random variables $(I_U^v)_{v\notin U}$ are independent, it is well known that the law of their sum, that is the law of  $N_U$, can be approximated by a Poisson law \citep{BHJ92}. More precisely, denoting by $d_{TV}$ the total variation distance between two distributions, we have 

$$ d_{TV}\big( \mathcal{L}(N_U) , \mathcal{L}(Po(\lambda_U))\big) \leq \min(1,\lambda_U^{-1}) \sum_{v\notin U} (p_U^v)^2, $$

where $\mathcal{L}(Po(\lambda_U))$ is the Poisson distribution with parameter $\lambda_U$.  

This approximation may be used to determine an upper bound for the $p$-value of testing if the $(\mot, C)$-theme order is surprisingly large. In practice, such bounds are quite accurate as the $p_U^v$'s are small. 

Nevertheless, a better approximation can be obtained for the tail probabilities by using Chen-Stein's method \citep{Chen75}, as shown in \citeauthor{BHJ92}. 
\begin{equation}
\forall K>2 \lambda_U,\quad
\proba(N_U\geq K| G[U] \sim \mot') \leq \frac{K-\lambda_U}{K-2\lambda_U} Po(\lambda_U)([K, +\infty)), \label{localeq1}
\end{equation} 

where, for any measurable set $A$, $Po(\lambda_U)(A)$ is the probability of $A$ under the Poisson distribution with parameter $\lambda_U$.

Setting $K=\lceil \lambda_U(1+t) \rceil$ for some $t>1$ and using elementary bounds and transformations developed in Appendix~\ref{appendixA}, we obtain
\begin{equation}
\forall t>1, \proba(\Delta_U \geq t) \leq  \proba(G[U] \sim \mot') \frac{\sqrt{t+1}}{\sqrt{2\pi \lambda_U} (t-1)} e^{-\lambda_U ((1+t)\ln(1+t) -t)} \label{localeq2}
\end{equation}

For positive values of $t$ which are smaller than or close to $1$, a sharper bound can be obtained by using a concentration inequality on the sum of independent random variables bounded between $0$ and $1$ \citep{McD98}.
\begin{equation}
\forall t>0,\quad 
\proba(\Delta_U\geq t) \leq \proba(G[U] \sim \mot') e^{-\lambda_U ((1+t)\ln(1+t) -t)}. \label{localeq3}
\end{equation}

Moreover, it is straightforward to verify that the function of $t$ defined on $]1,+\infty[$ by $\frac{\sqrt{t+1}}{\sqrt{2\pi \lambda} (t-1)}$ is decreasing  and is equal to $1$ at 

$$t_\lambda =1 + \frac{1}{4\pi \lambda} ( 1+ \sqrt{1+16\pi \lambda} ).$$

Therefore, coupling inequalities \eqref{localeq2} and \eqref{localeq3} yields a local 
bound for the tail probability of the theme order.

\begin{thm}\label{localthm}
For any pattern $\mot$, subpattern $(C,\mot')$ and position $U$ and for every positive $t$, let

$$
h(\lambda,t) = \left\{ \begin{array}{cl}
1 & \mathrm{ if } \quad t\leq t_{\lambda}, \\
\frac{\sqrt{t+1}}{\sqrt{2\pi \lambda} (t-1)} & \mathrm{ if } \quad  t>t_{\lambda}.
\end{array}
\right.
$$

Then, $\forall t>0$,

$$
\proba(\Delta_U\geq t)  \leq  \proba(G[U] \sim \mot') h(\lambda_U,t)  e^{-\lambda_U ((1+t)\ln(1+t) -t)}. \label{localeq4}
$$

\end{thm}

We thus have an exponentially decreasing local bound for the tail probability of the centered and renormalized order of an $(\mot, C)$-theme on $U$. Moreover, that bound is easily computable from  the parameters of the random  graph model.

\subsection{A global statistic to detect local motifs}

Theorem~\ref{localthm} allows to test whether there is a local over-representation at a given position $U$ of an $(m,C)$-theme. However, the number of possible positions $U$ is growing as $n^{k-1}$.  We thus encounter
a multiple testing problem. To overcome this issue, we build a statistic characterizing any local over-representation of a pattern somewhere in the graph.

Let us consider the function $g$ defined for every positive $\lambda$ and $t$ by
\begin{equation}\label{gdefinition}
g(\lambda,t) = \lambda \big( (1+t)\log(1+t) -t \big) - \log(h(,\lambda,t)). 
\end{equation}
 
For any positive $\lambda$, the function $g(\lambda,.)$ is  a one-to-one increasing function, mapping $]0,+\infty[$ to itself and  which is equivalent
to $\lambda t \log(t)$ as $t$ tends to infinity. 
Thus, the event $g(\lambda_U,\Delta_U)$ much larger than $1$ is equivalent to the event $\Delta_U$ much larger than $1$.


For any positive $t$, let us apply Theorem~\ref{localthm} to $y$ such that $g(\lambda_U,y)=t$. We then obtain 
\begin{equation*}
\forall t>0,  \qquad  \proba \big(g(\lambda_U,\Delta_U) \geq t  \big)\leq \proba(G[U]\sim \mot') e^{-t}. 
\end{equation*} 

Noting that
the event $E^t =\{ \max_U( g(\lambda_U,\Delta_U) ) \geq t\}$ is the union over all the possible positions $U$ of the events $E^t_U = \{g(\lambda_U,\Delta_U) \geq t \}$,  and that the 
exponential term in the upper bound is independent of $U$, we obtain our main result, stated in the following theorem.

\begin{thm}\label{main}

Let $g$ be the function defined in Equation~\eqref{gdefinition} and $N(\mot')$ the random variable denoting the global number of occurrences of $\mot'$ in $G$. Then, for every $t>0$, 
\begin{equation}
\proba \big(  \max_{U}(g(\lambda_U,\Delta_U))  \geq  t  \big)  \leq  \esp N(\mot')    e^{-t} \label{globaleq}
\end{equation} 

\end{thm}

We thus obtain an upper bound on the global $p$-value for detecting a local over-representation of $\mot$ with respect to the subpattern
$(C,\mot')$ occurring anywhere in the network.

\subsection{Motif selection criterion}\label{motifselection}

Consider the two patterns and respective subpatterns of Figure~\ref{themeexample}. Let $C_1$ and $C_2$ denote the respective deletion classes of the subpatterns. Then, as shown by the figure, every $(\mot_2,C_2)$-theme of order $K$ is an $(\mot_1,C_1)$-theme of order $K+1$.  In that case,  the fact that the $(\mot_2,C_2)$-theme is of order significantly larger than expected is redundant with the same information for the $(\mot_1,C_1)$-theme.

To avoid such redundance in the final motif list, a pattern $\mot$ will be considered as a motif with respect to a subpattern $(C, \mot')$ if the two following conditions hold:
\begin{enumerate}
\item The $p$-value given by Theorem~\ref{main} is lower than a fixed threshold, that is $\mot$ is a potential local motif;
\item Let $\{a\}$ be a vertex of $C$. There exist no set $\mathcal{A}$ of vertices of $\mot$ such that
\begin{itemize}
\item there is no edge between $a$ and any vertex of $\mathcal{A}$,
\item $\mot \setminus \mathcal{A}$ is over-represented with respect to $(D, \mot \setminus (\mathcal{A} \cup \{a\})$, where $D$ is the deletion class of $\{a\}$ in $\mot \setminus \mathcal{A}$ .
\end{itemize}
\end{enumerate}

\begin{figure} [htb] 
\psfrag{A}{$c$}
\psfrag{B}{$d$}
\psfrag{S}{$a$}
\psfrag{T}{$b$}
\psfrag{u}{$u$}
\psfrag{v}{$v$}
\psfrag{w}{$w$}
\psfrag{petita}{($1$)}
\psfrag{petitb}{($2$)}
\begin{center}
\epsfig{file=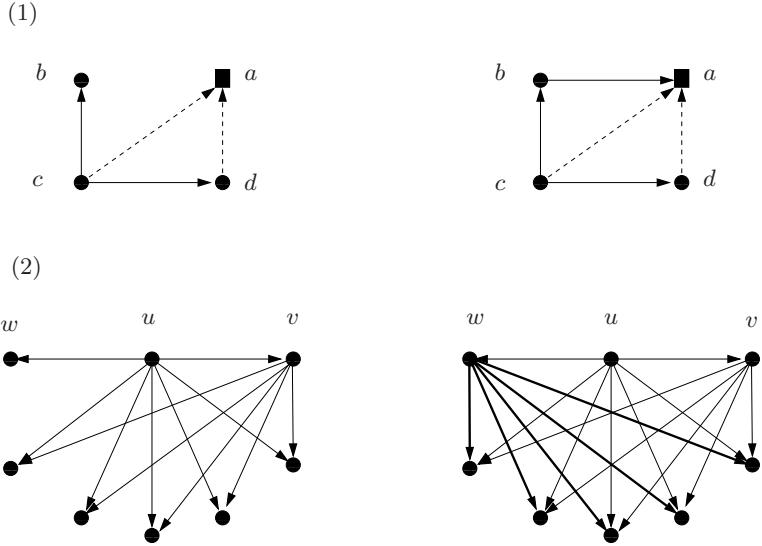, width=.8\textwidth}
\end{center}
\caption{Illustration of the filtering procedure. ($1$) Suppose the two patterns shown  are potential local motifs with respect to the deletion of $a$ and that the feed-forward loop was previously declared as a motif with respect to the deletion of $a$. Then the left pattern is filtered out by applying the filtering procedure for $\mathcal{A}=\{b\}$. However, the second is declared as a motif because of the edge between $b$ and $a$. ($2$) Themes of order $5$ in the network $G$ for the two previous potential motifs. In the second case, the edge from $b$ to $a$ in the pattern implies the presence of $5$ additionnal edges in the theme.}
\label{filteringfig}
\end{figure}

If there exists a set $\mathcal{A}$ satisfying the two points of the second condition, then the over-representation of $\mot$ with respect to $(C, \mot \setminus \{a\})$ is considered as redundant with the over-representation of $\mot \setminus \mathcal{A}$ with respect to $(D,\mot \setminus (\mathcal{A} \cup \{a\}))$. Thus, the pair $(\mot, C)$ is filtered out from the local motif list. 

Figure~\ref{filteringfig} illustrates the filtering procedure. Moreover, it shows why the absence of any edge between vertex $a$ and set $\mathcal{A}$ is required to filter out a potential local motif. Indeed, consider the first pattern of the figure for which $\mathcal{A}=\{b\}$ fulfills the previous condition. The corresponding theme doesn't convey any new information compared to the theme of the feed-forward loop obtained by deleting the vertex $w$. 

On the opposite, a theme of order $k$ of the second pattern  contains $k$ supplementary edges compared to the feed-forward loop theme.   
 The coefficients of $\mathbf{\Pi}$ being small in general because of the sparsity of real networks, the presence of those $k$ edges is informative. Thus, that potential local motif is kept in the list of local motifs.

\subsection{Algorithmic issues}

 Given an integer value $k$, our procedure first needs to list all the patterns of size $k$ occirring in the network, as well as their subpatterns. 
 That issue is tackled by using the ESU algorithm of \cite{Wer05}. 
 It is important to note that this subgraph count has only to be done on the observed graph and not on a huge number of simulated ones.  

We then apply Theorem~\ref{main} to each pair (pattern, subpattern).
The major cost in terms of computational time of that step is the computation of $\esp(N(\mot'))$. 
Indeed, it requires to sum a probability of occurrence along all possible positions in the graph.
It may be done more efficiently by grouping the nodes belonging to the same class. This approach gives good results when the number of classes is not too large, which is the case in practice when estimating the classes using mixture models algorithms. However, it becomes a real drawback for motifs larger than $4$ vertices when using the {\em Expected Degree} method and observed graph with more than $200$ vertices.

Finally, as the algorithm visits every position, it is not time-consuming to keep in memory all the positions and their respective theme orders in order to have a better interpretation of the results.

\section{Lower bound}\label{lowerboundsection}

  The fact that Theorem~\ref{main} gives an upper-bound of the exact $p$-value ensures that 
the number of false positives is controlled by the threshold used in the procedure. However, the tightness of that bound has to be taken into consideration to tackle the problem of false negatives.

An evaluation of the tightness of  the local bound given by Theorem~\ref{localthm} can be obtained for moderate deviations, as stated in the following proposition.

\begin{prop}\label{LWproplocal}
 Consider any pattern $\mot$ and subpattern $(C, \mot')$.
 Let $U$ be a position corresponding to an occurrence of $\mot'$ in $G$ and define $\lambda_{2,U}=\sum_{v\notin U} (p_U^v)^2$. 
Denote by $B_U(t)$ the local upper bound on $\proba(\Delta_U\geq t)$ given by Theorem~\ref{localthm}.

Suppose that $\lambda_{2,U}< \frac{1}{4}$.
Then, for every $t$ such that $1< t < \frac{1}{8\sqrt{\lambda_{2,U}}}-1$,  

$$\frac{\proba(\Delta_U\geq t)}{B_U(t)} \geq \Big[ 1-52 \frac{\lambda_{2,U}}{\lambda_U}(1+t)\Big] \Big[1 - \frac{2}{1+t}\Big] \Big[1- \frac{1}{10\lambda_U(1+t)} \Big] $$
 
\end{prop}

The proof relies on results about Poisson approximations for sums of independent random variables given in \citeauthor{BHJ92} and is detailed in Appendix~\ref{appendixB1}.

This theorem shows that for infinite sequences of real numbers $t^{(n)}$, graphs $G^{(n)}$ and positions $U^{(n)}$ such that $t^{(n)}$ and $(t \lambda_U)^{(n)}$ go to infinity and $(t\lambda_{2,U}/\lambda_U)^{(n)}$ goes to $0$, the bound on $\proba(\Delta_U^{(n)}\geq t^{(n)})$ is asymptotically tight. 

For example, let us consider the Erd\H{o}s-R\'enyi model with a connection probability $p^{(n)}=\frac{c}{n}$. That choice corresponds to a linear growth of the number of edges \citep{CL06}. Denote by $k$ and $n$ the respective sizes of $\mot$ and $G$ and by $r$ the number of edges which are in $\mot$ but not in $\mot'$. Then, for any position $U$,
\begin{eqnarray}
\lambda_U^{(n)} & = & (n-k+1) p^r  \quad \sim_{n\to +\infty}  \frac{c^r}{n^{r-1}} \\
\lambda_{2,U}^{(n)} & = & (n-k+1) p^{2r} \quad  \sim_{n\to +\infty}  \frac{c^{2r}}{n^{2r-1}}  
\end{eqnarray} 

Thus, for $r\geq 2$, choosing $t^{(n)} \sim n^{\alpha}$  with $r-1< \alpha<r$ yields a sequence of thresholds for which the bound of Theorem~\ref{localthm} is asymptotically tight.
\smallskip

The tightness of the global bound given by Theorem~\ref{main} is a more intricate issue. 
Using the notations $E^t =\{ \max_U( g(\lambda_U,\Delta_U) ) >t\}$ and $E^t_U = \{g(\lambda_U,\Delta_U)>t \}$,
the derivation of a lower bound for the event $E^t$ can be done using 

$$ \proba(E^t) \geq \sum_U \proba(E^t_U) - \sum_{U,V} \proba(E^t_U \cap E^t_V). $$

The term $\sum_U \proba(E^t_U)$ corresponding to the proposed upper bound, it is sufficient to derive tight upper bounds on the intersections $E^t_U \cap E^t_V$.  
Nevertheless, for some patterns, the number of extensions at two overlapping positions $U$ and $V$ may be strongly correlated. For instance, consider Figure~\ref{themeexample} and in particular the occurrence of the pattern $\mot_2$ at position (PDR$1$, PDR$3$, PDR$5$, IPT$1$). The number of extensions of $\mot'_2$ at position $U=$(PDR$1$, PDR$3$, PDR$5$) will be equal to the number of its extensions at position $V=$(PDR$1$, PDR$3$, IPT$1$). Therefore, the probability of the intersection is not small with respect to the probabilities of the single events. 

However,  this approach allows us to show that the global upper bound we propose is tight in the sense that for some patterns and some random models corresponding to sparse graphs, it is asymptotically the best one.

\begin{prop}\label{LWpropglobal}
Let $\mot$ be a pattern of size $k$ admitting some vertex $a$  linked to every other vertex of $\mot$. Consider its subpattern $(C, \mot')$ where $C$ is the deletion class of $a$.

Let $\rho = \max_{i,j} \Pi_{i,j}$ and suppose that $\rho=\mathcal{O}(n^{-\frac{1}{2}-\epsilon})$, with $\epsilon>\frac{1}{2k}$. Let $\delta=min(\epsilon, 2k\epsilon -1)>0$. Then
$$\proba(\max_{U}(g(\lambda_U,\Delta_U))  > t)  = (1-\eta)\sum_{U} \proba(g(\lambda_U,\Delta_U)  > t) , \quad \mbox{where }\eta = \mathcal{O}(n^{-\delta})$$
\end{prop}

Note that the condition on $\rho$ still allows a growth rate of the number of edges of the order  $\mathcal{O}(n^{\frac{3}{2}-\epsilon})$, which is faster than the linear growth observed on real data \citep{CL06}. The detailed proof of the proposition is given in Appendix~\ref{appendixB2}.

Combining Propositions~\ref{LWproplocal} and \ref{LWpropglobal} yields that the proposed upper bound is asymptotically tight for some pairs of patterns and subpatterns in a given range of model parameters.


\section{Illustration of the procedure}\label{results}

\subsection{Simulated Data}

\begin{figure} [htb] 
\begin{center}
\epsfig{file=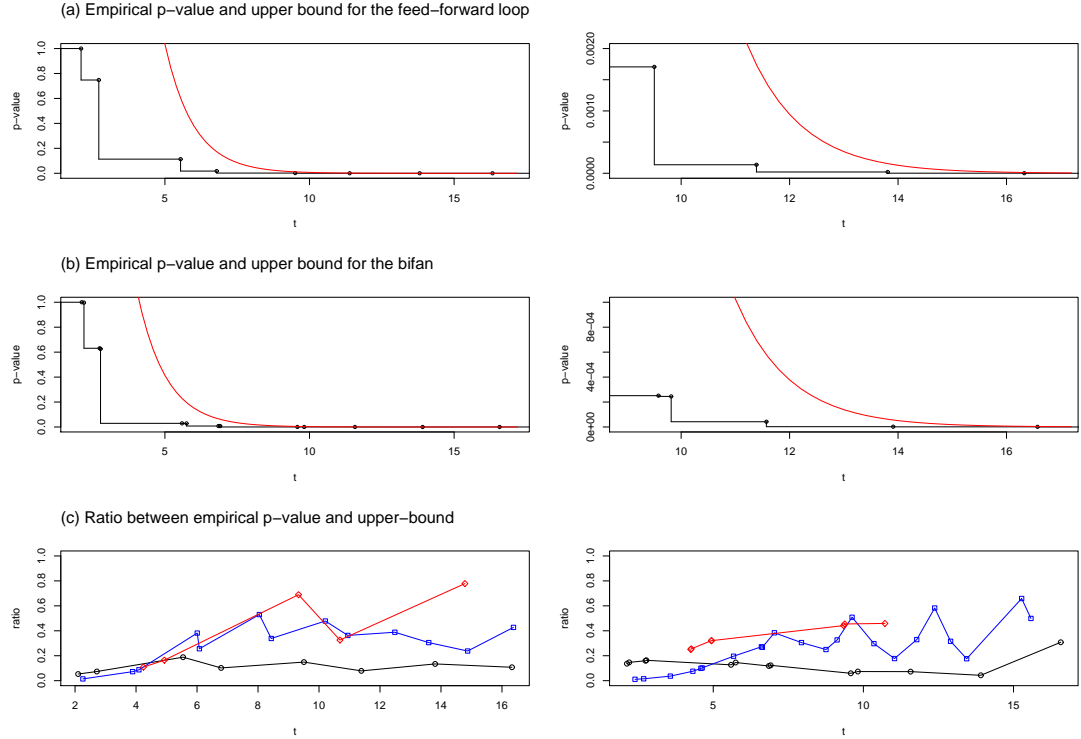, width=.8\textwidth, angle=-90}
\end{center}
\caption{Empirical $p$-values and corresponding upper bound. (a) Both curves for the feed-forward loop in the 500,000 reference graphs and zoom at the distribution tail. (b) Same as (a) for the bi-fan pattern. (c) Ratio between the empirical $p$-value and the upper bound for the reference graphs (black circles), the dense graphs (blue squares) and the large graphs (red diamonds).}
\label{simulationfig}
\end{figure}

$500,000$ directed graphs with $90$ vertices, which we will call the {\em reference graphs},  were generated under the model with three classes of $30$ vertices each and connection probabilities set to $0.04$ between vertices of the same class and $0.01$ between vertices of different classes.
 The mean out-degree and in-degree under that model are both equal to $1.76$.
 Our method is illustrated with  the feed-forward loop an the bi-fan patterns (see Figure~\ref{delclasses}).
 The choice of the subpatterns is done by deleting the only vertex of in-degree $2$ for the feed-forward loop and one of the vertices of in-degree $2$ for the bi-fan. Nevertheless, that choice plays no role in that particular case, due to the symmetry of the model. 
Figure~\ref{simulationfig} (a) and (b) show the empirical tail probabilities  and corresponding upper bounds given by Theorem~\ref{main}, as functions of the parameter $t$. 

To evaluate the quality of the upper bound, the ratio between the empirical $p$-values and their upper bounds is shown in Figure~\ref{simulationfig} (c). The same ratio is also plotted for more dense graphs ($30,000$ graphs sampled with the same number of nodes and classes and connection probabilities five times larger) and for graphs larger than the reference graphs but with comparable density ($30,000$ graphs with $360$ nodes, three classes of $120$ nodes each and connection probabilities $0.01$ and $0.0025$). 
For the reference graphs, the ratio is about $1/10$ for both patterns, with a minimum of $0.07$ for the feed-forward loop and $0.04$ for the bi-fan. This ratio increases both for larger graphs and more dense graphs.

\subsection{Stability with respect to the random model estimation}

\begin{table}[ht]
\begin{center}
\begin{tabular}{|c|c|c|c|c|}
\hline
          &  Erd\H{o}s & ED & MixNet & BLOCKS   \\
\hline
& & & & \\
\epsfig{file=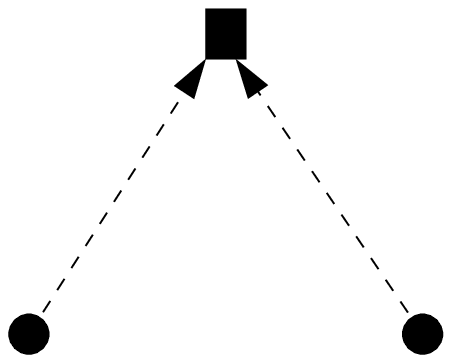, height= .6cm}
& 4.3 e-46 &  & 5.7 e-7 & 1.3 e-6  \\
\hline
& & & & \\
\epsfig{file=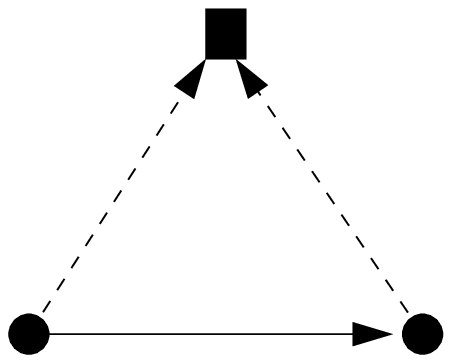, height= .6cm}
& 5.2 e-14 & 5.8 e-4 & 8.2 e-6 & 3.9 e-5  \\
\hline
& & & & \\
\epsfig{file=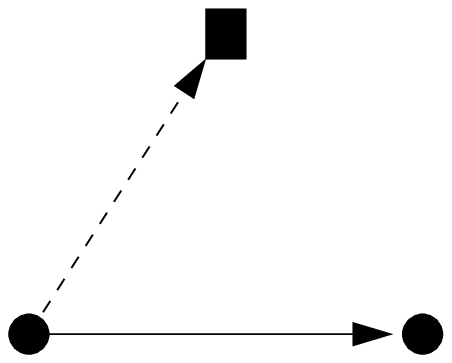, width=.8cm}
& 1.3 e-30 &  & & \\
\hline
& & & & \\
\epsfig{file=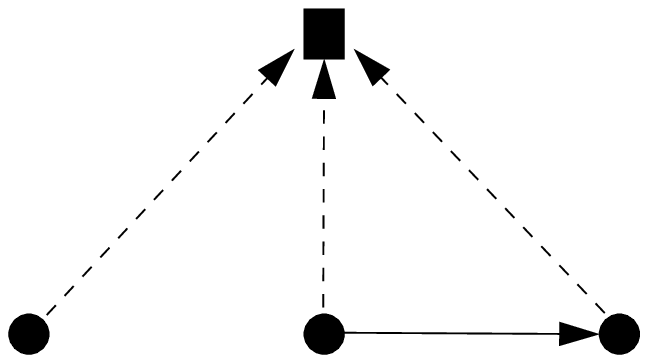, width=1cm}
&  3.2 e-14 & 2.9 e-5 & 3.1 e-8 & 9.6 e-5 \\
\hline
& & & & \\
\epsfig{file=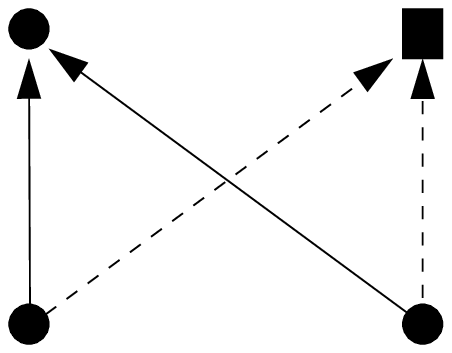, width=.8cm}
&  (4.6 e-45) & 9.6 e-4 & (4.1 e-9) & (1.2 e-6) \\
\hline
& & & & \\
\epsfig{file=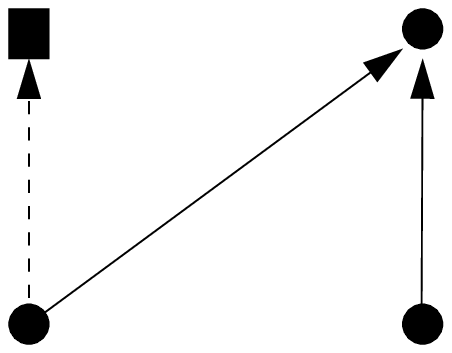, width=.8cm}
&  (1.9 e-30) & & 3.5 e-2  & 3.8 e-2   \\
\hline
& & & & \\
\epsfig{file=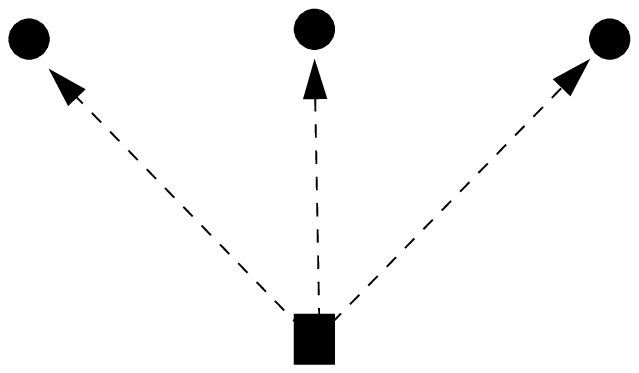, width=1cm}
& 1.6 e-4 & &   &   \\
\hline
\end{tabular}
\end{center}
\caption{Local motifs found with different estimation procedures for the parameters of the random model. The table gives the upper-bound on the $p$-value computed by \eqref{globaleq}. The values under brackets show potential motifs which are filtered out as redundant with a smaller motif.}
\label{stabilitytable}  
\end{table}      

The $p$-value used to decide if a pattern is a motif or not depends on the inferred connection matrix $\mathbf{\Pi}$. 
To evaluate the incidence of the choice of the inference method, the motif search is run for four distinct models:
\begin{itemize}
\item the Erd\H{o}s-R\'enyi model \citep{ER59} with a connection probability such that the expected number of edges is equal to  the observed one;
\item the {\em Expected Degree} ({\em ED}) model described in \citeauthor{MSB06}. That model draws a link between vertices $u$ and $v$ with a probability proportional to $d_ud_v$, where $d_u$ denotes the observed degree of the node $u$. An adequate choice of the normalisation constant leads to a model for which the expected degree of each node is almost equal to its observed one;
\item the mixture model {\em MixNet} in its Bayesian version \citep{LBA08} implemented in the {\em mixer} R-package;
\item the mixture model {\em BLOCKS} \citep{NS01} available in the {\em STOCNET} software ({\em http://stat.gamma.rug.nl/stocnet/}).
\end{itemize}

As {\em BLOCKS} only supports graphs  up to $200$ nodes, we consider a randomly chosen subnetwork of the transcriptional Yeast regulation network containing $194$ nodes. Inference of the matrix $\mathbf{\Pi}$  and  search for all local motifs of
 size $3$ and $4$ is done on that subnetwork.  
Table~\ref{stabilitytable} shows all the motifs found using a threshold of $0.05$ on the $p$-value with at least one of the method.  One can see that the list of motifs remains stable. 

The only method leading to significantly different $p$-values is the one relying on the Erd\H{o}s-R\'enyi model, 
which is known to poorly describe real networks.

On the other hand, the method relying on the {\em Expected Degree} model is the one selecting the smallest number of motifs.The main difference with the two last methods is that the top-motif for {\em Mixnet} and {\em BLOCKS} is not a motif for {\em ED}. However, large themes for the bi-fan, which is detected as a motif  by {\em ED}, correspond also to large themes of the first motif of size $3$. Thus, the themes pointed out by the three methods are the same ones.  

Finally, the two methods taking into account both the degree-distribution and the group structure of networks, that is {\em MixNet} and {\em BLOCKS}, select the same motifs and with comparable $p$-values.

\subsection{Real networks}

In order to point out the difference between global motifs and local ones, we run  
our  procedure to determine all local motifs of size $3$, $4$ and $5$ in two standard networks, both studied in \citeauthor{MS02}, and publicly available at {\em http://weizmann.ac.il/mcb/UriAlon}. Those networks are the transcriptional regulatory networks of Yeast  and the electronic circuit {\em s420} of the {\em ISCAS89} benchmark.  

Inference of the parameters of the model is done using the Bayesian MixNet approach in order to be able to tackle patterns of size $5$ in those graphs. However, that choice implies that hubs may be grouped in the same class, which is relevant from a mixture model point of view  but may generate quite inhomogeneous groups in terms of degrees and thus local motifs with poor biological interpretation. 
For example, the parameter inference on the Yeast network gives rise to a group of two hubs of respective out-degrees $71$ and $44$.  Hence, the expected outdegree in that group is $57.5$ and any star pattern will be selected as a local motif when centering the position on the largest hub. 

To avoid that phenomenon, we first run the whole procedure with the Bayesian MixNet approach and run it again with the Expected Degree approach when   
the position of the theme leading to a local motif shows that it is selected because of the presence of a vertex of high degree.


 \begin{table}[ht]
\begin{center}
\begin{tabular}{|c|c|c|c|c|}
\hline
& & & & \\
  Local motif  & \epsfig{file=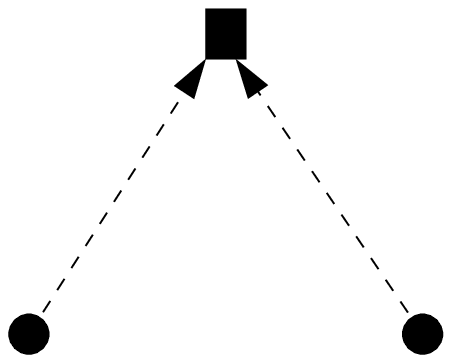, height= 1cm} & \epsfig{file=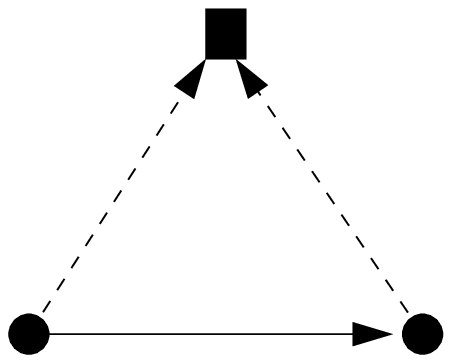, height= 1cm} & \epsfig{file=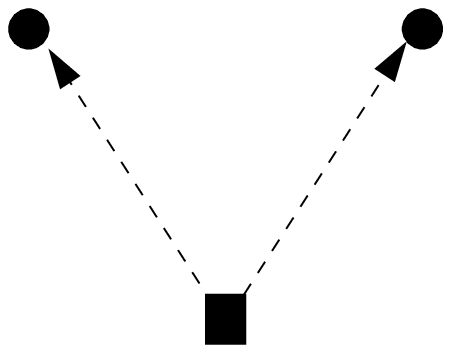, height= 1cm} & \epsfig{file=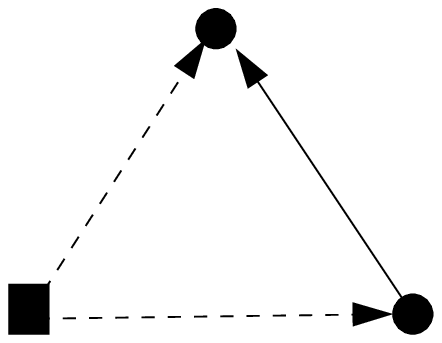, width=1.2cm} \\
& & & & \\
\hline
$p$-value bound & 2.0 e-16 & 2.3 e-9 &  4.6 e-4 & 8.6 e-4 \\
\hline
 $N_U^*$  & 38 & 15 & 5 & 3   \\
\hline
\end{tabular}
\end{center}
\caption{Local motifs  of size $3$ found in the Yeast regulatory network.}
\label{yeast3table}  
\end{table}

The algorithm is run on the Yeast regulatory network with a threshold of $1 e-3$. 
Table~\ref{yeast3table} shows the local motifs of size $3$ found in the network, the row $N_U^*$ denoting the order of the theme at the position $U$  where $\Delta_U$ is maximal. 

 There are clearly two top local motifs of size $3$. 

The first top local motif corresponds to a pair of regulators co-regulating a gene. This motif is not selected by the methods of global motif detection of \citeauthor{MS02}, \citeauthor{BL04} or \citeauthor{WR06}. However, all those methods select the motif of size $4$ called bi-fan (see Figure~\ref{delclasses}). That global over-representation of  the bi-fan is a consequence of the the local motif we detect. Indeed, the three largest themes of our local motif  are of respective orders $38$, $32$ and $18$. Thus, their presence imply ${38 \choose 2} + {32 \choose 2} + {18 \choose 2}= 1352$ occurrences of the bi-fan. As the total number of occurrences of the bi-fan in the network is $1843$, it gives confirmation on the local character of the over-representation of that pattern.

 The second one is the feed-forward loop with respect to the deletion of the vertex in-degree $2$. A feed-forward is composed by a main regulator $X$ and a gene $Y$ regulated by $X$, both co-regulating a third gene $Z$. This pattern is a local motif with respect to the subpattern obtained by deleting $Z$. This indicates the existence of places in the network where a main regulator $X$ and a gene $Y$ regulated by $X$ both co-regulate a high number of genes $Z_1,\ldots ,Z_k$. The value of $N_U^*$ indicates that there is at least such a theme of order $38$. That phenomenon is already described by U. Alon \citep{Al07}, under the denomination {\em Multi-output feed-forward loops}. 

The feed-forward loop is also found to be a local motif  with respect to the deletion of the main regulator $X$. However,  the $p$-value upper bound indicates that the order of the corresponding themes is less significative than in the previous case. This fact is confirmed by the lower value of $N_U^*$. It illustrates a main point of our method which is to differentiate the behaviour of the different deletion classes of a pattern.

Finally, the feed-forward loop is not a local motif with respect to the deletion of the intermediate gene $Y$, indicating that no main regulator  $X$ regulates a high number of genes $Y_1\ldots Y_k$ in order  to regulate a gene $Z$.


\begin{table}[ht]
\begin{center}
\begin{tabular}{|c|c|c|c|c|c|}
\hline
& & & & &  \\
 Local motif   & \epsfig{file=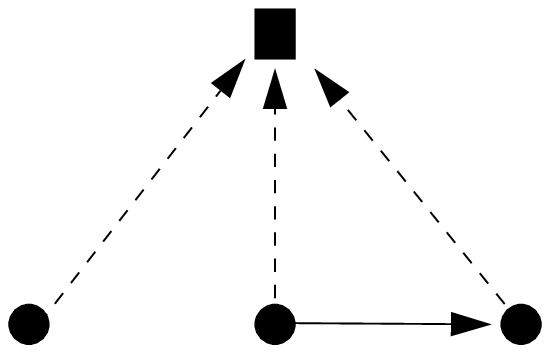, height= .9cm} & \epsfig{file=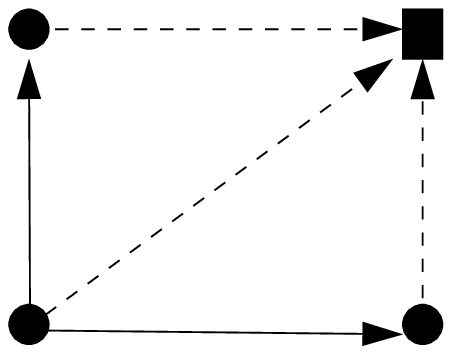, width=1cm}  & \epsfig{file=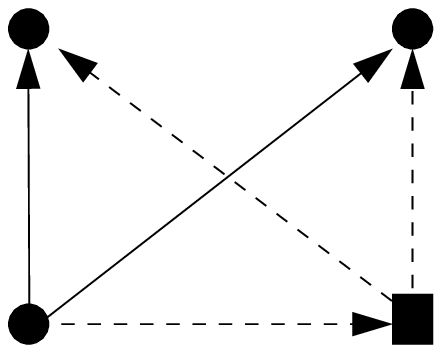, width=1cm} & \epsfig{file=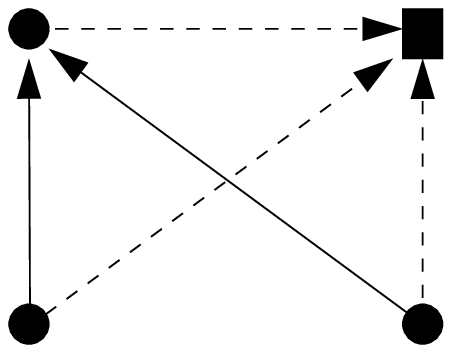, width=1cm}  & \epsfig{file=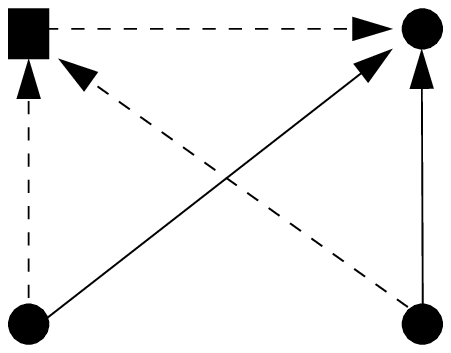, width=1cm}\\
& & & & & \\
\hline
$p$-value bound & 6.5 e-15 & 3.4 e-6  & 1.4 e-4  & 5.6 e-4 & 9.2 e-4 \\
\hline
 $N_U^*$  & 7 & 2  & 2 &  1 & 1   \\
\hline
\end{tabular}
\end{center}
\caption{Local motifs found of size $4$ in the Yeast regulatory network.}
\label{yeast4table}  
\end{table}

The method finds five local motifs of size $4$ in the Yeast regulatory network. They are shown in Table~\ref{yeast4table}. 

The first motif appearing in the list is of interest as it corresponds to three regulators co-regulating seven genes with an additional regulation between two of them. Another way to see the theme of that motif is a multi-output feed forward loop of order $7$ with a third regulator acting on $Z_1,\ldots, Z_7$. That motif is also found by the global methods but with a higher $p$-value and at the fifth or sixth position among the motifs of size $4$.

The $4$ other local  motifs have a value of $N_U^*$ lower than $2$, suggesting that they appear in the list because of a very low expected value. Note that the bi-fan does not appear in the list as it is filtered out as a redundancy of the second motif of size $3$.

Finally, there is only one local motif of size $5$. It corresponds to the fourth motif of size $4$ with a supplementary edge going out from the squared vertex. Its value for $N_U^*$ is $1$, showing that its expected is low, such that it becomes over-represented even at its first occurence. 
\smallskip

The second network, that is the electronic circuit, has no local motif of size $3$, $4$ or $5$ relying on a threshold of $.01$.  However, one global motif of size $3$ and two global motifs of size $4$ were found in \citeauthor{MS02} with $Z$-scores larger than $10$. This indicates a distinct behaviour of the two types of networks, the occurrences of the global  motifs of the electronic network being spread in the whole networks rather then agglomerated.


\section{Conclusion}\label{conclusion}

 In this work, we propose a new approach to study network motifs, that is to look for locally over-represented patterns. Our framework allows us to take into account the over-representation of a pattern with respect to its subpatterns, for any pattern size. To list the local motifs of a network, we use a model-driven approach to determine a $p$-value upper bound for each pair (pattern, subpattern) and then apply a filtering procedure to eliminate redundancy. 
 
Simulated data show that the error made by taking an upper-bound of the exact $p$-value is reasonable. The application of our method on standard real data allows us to find information on the role of the vertices of the motifs only by statistical means. Moreover, comparing the lists of local and global motifs highlights a strong structural difference between networks of different nature. In future work, we will investigate non standard data and  a deeper understanding of the local motifs which are not global ones.

\section*{Acknowledgements}

The author would like to thank Catherine Matias and Gesine Reinert for their remarks and suggestions and  Gilles Grasseau for his help while implementing the method.


\bibliographystyle{imsart-nameyear}      
\bibliography{References}

\begin{appendix}

\section{Local upper bound}\label{appendixA}

To prove Inequality~\eqref{localeq2}, we start from Inequality~\eqref{localeq1} which corresponds to Theorem 2.R in \citeauthor{BHJ92} and apply it for $K= \lceil \lambda_U(1+t) \rceil$. Then, $\Delta_U\geq t$ if and only if $N_U\geq K$.

For all $k\geq K$, let $u_k=\frac{\lambda_U^k}{k!} e^{-\lambda_U}$.
Then 
$Po(\lambda_U)([K, +\infty)) = \sum_{k\geq K} u_k $ 
and 
$\forall k\geq K, \frac{u_{k+1}}{u_k} \leq \frac{1}{t+1} $.

Thus, using that $K! \geq \sqrt{2\pi K}(\frac{K}{e})^K$, we get
\begin{eqnarray*}
Po(\lambda_U)([K, +\infty)) & \leq & \frac{1}{1-\frac{1}{1+t}} \frac{\lambda_U^K}{K!} e^{-\lambda_U}   \\
    & \leq & \frac{t+1}{t}  \frac{\lambda_U^K e^K}{\sqrt{2\pi K} K^K} e^{-\lambda_U} 
\end{eqnarray*}

Then, using Inequality~\eqref{localeq1},
\begin{eqnarray*}
\proba(\Delta_U\geq t| G[U] \sim \mot') 
 & \leq & \frac{t}{t+1}  \frac{t+1}{t \sqrt{2\pi \lambda_U (1+t)}}  \frac{\lambda_U^{\lambda_U(1+t)} e^{\lambda_U(1+t)}}{ (\lambda_U(1+t))^{\lambda_U(1+t)}} e^{-\lambda_U} \\
 & \leq & \frac{1}{\sqrt{2\pi \lambda_U (1+t)}} e^{-\lambda_U((1+t)\log(1+t) -t)}                       
\end{eqnarray*}

Writing that 
$$\proba(\Delta_U\geq t) =  \proba(\Delta_U\geq t| G[U] \sim \mot') \proba(G[U] \sim \mot')   $$
 yields Inequality~\eqref{localeq2}.

\section{Lower bound}

\subsection{Local lower bound}\label{appendixB1}

The first step is to find the best possible bound for the difference between the tail probability of a sum of independent random variables and the tail probability of the corresponding Poisson approximation. This problem is presented and studied in~\cite{BHJ92}. We use Theorem 9.D presented in that book, namely
\begin{thm}[Barbour, Holst, Janson]\label{BHJthm}
Define $W=\sum_i X_i$, where $X_i$ are independant random variables. Set $\lambda=\sum_i \esp(X_i)$ and $\lambda_2=\sum_i \esp(X_i)^2$.

Let $K\geq \lambda$ be an integer, $\xi=\lambda_2 / \lambda$ and $\Gamma=(K-\lambda)/\sqrt{\lambda}$. Then, uniformly in $K$ satisfying $\Gamma\geq 1$, $K\leq \lambda / 2\xi$ and $1+4\Gamma^2\leq (16\xi)^{-1}$, we have

$$ \proba(W\geq K) = Po(\lambda)([K,+\infty)) (1+\mathcal{O}(\xi)+\mathcal{O}(\xi \Gamma^2)).$$
\end{thm}

Applying this result in our context for $W=N_U$ and $K=\lceil \lambda_U(1+t) \rceil$ for a fixed $t$ may not be possible because in this case $\Gamma=t\sqrt{\lambda_U}$ and thus the condition $\Gamma\geq 1$ may not be satisfied when $\lambda_U$ is too small with respect to $t$.

However, the proof of Theorem~\ref{BHJthm} uses the assumption $\Gamma\geq 1$ only once. Rewriting it without that condition until that step yields:

$$ \frac{\proba(W\geq K)}{Po(\lambda)([K,+\infty))} =  (1+\eta_1)(1+\eta_2)$$

with 

\begin{equation} \label{LWlocaleta1}
|\eta_1|\leq \big(1+2\Gamma \lambda^{-1/2}\big)^2 2\lambda_2 / (K-\lambda) 
\end{equation}

 and 

\begin{equation}\label{LWlocalineq1}
 |\eta_2| \leq \sum_{r\geq K}  Po(\lambda)(r) |\epsilon_r| / Po(\lambda)([K,+\infty)) 
\end{equation}

where $|\epsilon_r| \leq 8\xi \Gamma^2 + 2(r-K) \xi \Gamma \lambda^{-1/2}$.

The hypothesis $\Gamma\geq 1$ is then used to bound the right hand side of Inequality~\eqref{LWlocalineq1}, which can be alternatively bounded by
\begin{eqnarray*}
|\eta_2| & \leq & 8\xi \Gamma^2 + 2\xi \Gamma \lambda^{-1/2} \sum_{r\geq K} (r-K)\frac{Po(\lambda)(\{r\})}{Po(\lambda)\{[K,+\infty)\}} \\
         & \leq & 8\xi \Gamma^2 + 2\xi \Gamma \lambda^{-1/2} \sum_{r\geq K} r \frac{Po(\lambda)(\{r\})}{Po(\lambda)(\{K\})} \\
         & \leq & 8\xi \Gamma^2 + 2\xi \Gamma \lambda^{-1/2} \sum_{r\geq K} r \lambda^{r-K} \frac{K!}{r!} \\
         & \leq & 8\xi \Gamma^2 + 2\xi \Gamma \lambda^{-1/2} \frac{1}{1-\lambda/K},
\end{eqnarray*}

the last inequality deriving from the fact that the ratio between two consecutive terms of the sum is always lower than $\lambda/K$. 

Using the additional condition of Proposition~\ref{LWproplocal}, that is $K\geq 2\lambda_U$, and the fact that it implies $\Gamma \lambda^{-1/2}\geq 1$ allows us, using elementary bounds, to obtain from Inequalities~\eqref{LWlocaleta1} and \eqref{LWlocalineq1} that

\begin{equation}\label{LWlocalineq2}
|\eta_1| \leq 36 \xi \Gamma^2  \quad   \mbox{ and } \quad |\eta_2| \leq 16 \xi \Gamma^2. 
\end{equation}

As $\xi \Gamma^2 \leq \frac{K \lambda_{2}}{\lambda^2}$,  we have

$$ \proba(W\geq K) \geq Po(\lambda)\{[K,+\infty)\} (1- 52 \frac{K \lambda_{2}}{\lambda^2}). $$

The second part of the right hand-side of Proposition~\ref{LWproplocal} comes from the asymptotic comparison between the tail probability $Po(\lambda)\{[\lambda_U(1+t),+\infty)\}$  and its exponential approximation used in Theorem~\ref{localthm}. It is derived in a very similar way than in Appendix~\ref{appendixA}, using the following lower bound on 
 $K!$, which can be proved from its asymptotic expansion
$$ K! \geq \sqrt{2\pi K}\big(\frac{K}{e}\big)^{K}(1+\frac{1}{10K}). $$

\subsection{Global lower bound}\label{appendixB2}

\begin{lemma}\label{appendixLWlemma}
Let $\rho = \max_{i,j} \Pi_{i,j}$ and suppose that $\rho=\mathcal{O}(n^{-\frac{1}{2}-\epsilon})$, with $\epsilon>\frac{1}{2k}$.
 Let $\delta=min(\epsilon, 2k\epsilon -1)>0$. Then
$$\proba(E^t)  = (1-\eta)\sum_{U} \proba(E^t_U) , \qquad \mbox{where } \eta = \mathcal{O}(n^{-\delta}).$$
\end{lemma}

{\bf Proof:} We start from the simplest known lower bound for the probability of an union of events, that is:
\begin{equation}
\proba(E^t) \geq \sum_U \proba(E^t_U) - \frac{1}{2}\sum_{U} \sum_{V} \proba(E^t_U \cap E^t_V)
\label{appendixLWeq1}
\end{equation}

Given two positions $U$ and $V$, let $K_U$ and $K_V$ be such that $E^t_U = \{\Num \geq K_U\}$ and $E^t_V = \{N_V(\mot) \geq K_V\}$.
Moreover, let $ext(U)$ be the set of vertices yielding to extensions of $\mot'$ on $U$ and $i=|V\setminus U|$. 
Let us also recall that $k$ is the  size of the pattern $\mot$.

 Let $S$ be any set of vertices not intersecting $U$. We decompose $E^t_V$  as the union of the sets $\{T\subset ext(V)\}$
for all sets $T$ of $K_V$ vertices. Thus 
\begin{eqnarray*}
 \proba(E^t_V | ext(U) =S) & \leq & \sum_{T,|T|=K_V} \proba( T\subset ext(V) | ext(U) =S) \\
                       & \leq & \sum_{j=0}^{|S|}\quad \sum_{|T|=K_V, |T\cap S|=j} \proba( T\subset ext(V) | ext(U) =S). 
\end{eqnarray*}

Let $T$ be such that $|T\cap S|=j$. 
As $\mot'$ is connected, at least $i$ edges need to be present in $V\setminus U$ to ensure that $G[V]\sim \mot'$. 
Moreover, to ensure that $T\subset ext(V)$, all the edges between $V\setminus U$ and $T$ and all the edges between $U\cap V$ and $T\setminus S$ have to be present, which amounts to a total of at least $iK_V + (k-i)(K_V-j)$ edges.

Therefore, denoting by $\rho$ the largest coefficient of the matrix $\mathbf{\Pi}$,
$ \proba( T\subset ext(V) | ext(U) =S)\leq \rho^i \rho^{iK_V + (k-i)(K_V-j)}$.

We fix some $\epsilon>0$ and distinguish between two different cases whether the cardinality of $S$ is larger or smaller than $n^\epsilon$.

\begin{itemize}
\item If $|S|\leq n^{\epsilon}$, then
\begin{eqnarray}
 \proba(E^t_V | ext(U) =S) & \leq & \sum_{j=0}^{min(|S|,K_V)} {n \choose K_V-j} {|S| \choose j} \rho^{i+iK_V+(k-i)(K_V-j)}\nonumber \\
                       & \leq & \rho^{i+kK_V} n^{K_V} \sum_{j=0}^{K_V} {|S| \choose j} \big( \frac{1}{n\rho^{k-i}} \big)^j \nonumber\\
                       & \leq & \rho^{i+kK_V} n^{K_V} (K_V+1) \max (1, \big(\frac{1}{n\rho^{k-i}} \big)^{K_V}) \nonumber\\
                       & \leq & \rho^i (K_V+1) \max \big( n\rho^k, \rho^i |S|\big)^{K_V} \label{lemmaineq2}
\end{eqnarray}

It is straightforward to check that, for large enough $n$, we have $n\rho^k < \frac{1}{e}$ and $\rho^i |S|< \frac{1}{e}$. Therefore, the right hand side  of Inequality~\eqref{lemmaineq2} is a decreasing function of $K_V$ and 
\begin{equation}
\proba(E^t_V | ext(U) =S) \leq 2 \max \big( n\rho^{i+k}, \rho^{2i} |S|\big) \nonumber
\end{equation}

Moreover, $1-(i+k)(\frac{1}{2}+\epsilon) \leq -i-(2k\epsilon -1)$ for $\epsilon\leq \frac{1}{2}$ and $-2i(\frac{1}{2}+\epsilon)+\epsilon = -i-\epsilon$. Thus, as $\rho \leq C n^{-\frac{1}{2}-\epsilon}$ for some constant $C\geq 1$, and $|S|\leq n^{\epsilon}$,

\begin{equation}
\proba(E^t_V | ext(U) =S) \leq C^{2k} n^{-i-\delta}
\end{equation}\label{lemmaineq2bis}

\item For $|S|\geq n^{\epsilon}$, we roughly bound $\proba(E^t_V | ext(U) =S)$ by $1$. 
However, let us note that $g(\lambda_U,n^{\epsilon}) > n^{\epsilon}$ for large enough $n$ and therefore
\begin{eqnarray}
\sum_{S,|S|\geq n^{\epsilon}} \proba(ext(U)=S) & = & \proba(N_U \geq n^{\epsilon}) \\
                       & = & \proba(g(\lambda_U,N_U) \geq g(\lambda_U,n^{\epsilon}))\nonumber\\
                       & \leq &  \proba(g(\lambda_U,N_U) \geq n^{\epsilon})\nonumber\\
                       & \leq & \proba (E^t_U) e^{-n^{\epsilon}+t}\quad \mbox{ by Theorem~\ref{main}} \\
                       & \leq & \proba (E^t_U) n^{-i-\delta} \quad \mbox{ for large enough } n. \label{lemmaineq3}
\end{eqnarray}

\end{itemize}  

Using Inequalities~\eqref{lemmaineq2bis} and \eqref{lemmaineq3}, we get
\begin{eqnarray*}
\proba(E^t_U \cap E^t_V) & = & \sum_{S;|S|\geq K_U}\proba(E^t_V | ext(U) =S) \proba(ext(U)=S)\\
                     & \leq &  \sum_{K_U\leq |S|\leq n^{\epsilon}}C^{2k} n^{-i-\delta} \proba(ext(U)=S) + n^{-i-\delta} \proba (E^t_U)\\
                     & \leq & (C^{2k} +1) n^{-i-\delta} \proba (E^t_U).
\end{eqnarray*}

As the number of positions $V$ such that $|V\setminus U|=i$ is bounded by $n^i$, we finally obtain

\begin{eqnarray}
 \frac{\sum_{V\neq U} \proba(E^t_U\cap E^t_V)}{\proba(E^t_U)} & \leq & \sum_{i=1}^{k} n^i (C^{2k} +1) n^{-i-\delta} \nonumber \\
                    & \leq & k (C^{2k} +1) n^{-\delta}. \label{appendixLWeq2}
\end{eqnarray}

Inequalities~\eqref{appendixLWeq1} and \eqref{appendixLWeq2} yield the lemma.

\end{appendix}

\end{document}